\documentclass[12pt,preprint]{aastex}

\shorttitle{HV 11423}
\shortauthors{Massey et al.}

\begin{document}

\title{HV 11423: The Coolest Supergiant in the SMC}

\author{Philip Massey,\altaffilmark{1,2}
Emily M. Levesque,\altaffilmark{2,3}
K. A. G. Olsen,\altaffilmark{4}
Bertrand Plez,\altaffilmark{5}
B. A. Skiff \altaffilmark{1} }

\altaffiltext{1}{Lowell Observatory, 1400 W. Mars Hill Rd., 
Flagstaff, AZ 86001; Phil.Massey@lowell.edu; bas@lowell.edu.}

\altaffiltext{2} {Visiting Astronomer, Cerro Tololo Inter-American Observatory,
National Optical 
Astronomy Observatory (NOAO), which is operated by the Association of
Universities for Research in Astronomy (AURA), Inc., under cooperative agreement
with the National Science Foundation (NSF).}

\altaffiltext{3} {Institute for Astronomy, University of Hawaii at Manoa, 2680  Woodlawn Drive, Honolulu, HI 96822; emsque@ifa.hawaii.edu}

\altaffiltext{4} {Cerro Tololo Inter-American Observatory, NOAO, Casilla 603, 
La Serena, Chile; kolsen@noao.edu}

\altaffiltext{5}{GRAAL, Universit\'{e} de Montpellier II, CNRS, 34095 Montpellier, France; bertrand.plez@graal.univ-montp2.fr.}

\begin{abstract}

We call attention to the fact that
one of the brightest red supergiants in the SMC has recently changed its spectral
type from K0-1~I (December 2004) to M4~I (December 2005) 
and back to K0-1~I (September 2006). 
An archival spectrum from the Very Large Telescope reveals that the star
was even cooler (M4.5-M5~I) in December 2001.  By contrast, the star was
observed to be an 
M0~I in both October 1978 and October 1979. The M4-5~I spectral
types is by far
the latest type seen for an SMC supergiant, and its temperature
in that state places it well beyond the Hayashi limit into a region of the H-R diagram
where the star should not be in hydrostatic equilibrium. 
The star is variable by nearly 2 mag in $V$, but essentially
constant in $K$.  Our modeling of its spectral energy distribution shows that the visual
extinction has varied during this time, but that the star has remained  
essentially constant in bolometric
luminosity.  We suggest that the star is currently undergoing a period of intense
instability, with its effective temperature changing from 4300~K to 3300~K on the
time-scale of months.   It has one of the highest 12$\mu$m fluxes
of any RSG in the SMC, and we suggest that the variability at $V$ is due primarily
to changes
in effective temperature, and secondly, due to changes in the local extinction due to creation
and dissipation of circumstellar dust.   We speculate that the star may be nearing
the end of its life.

\end{abstract}


\section{Introduction}
\label{Sec-intro}

Red supergiants (RSGs) are the He-burning descendants of 10-25$\cal M_\odot$
stars.  They are not the most massive or luminous stars (those become
luminous blue variables and/or Wolf-Rayet stars instead), but their cool effective temperatures
and relatively high luminosities result in their being the physically largest
stars.   These stars are fully convective, with the lowest temperatures determined by
the Hayashi limit; cooler than this, stars are not hydrostatically stable.  The coolest
temperature allowed at a given luminosity should be somewhat higher for stars with
lower chemical abundances, as  in the Magellanic Clouds.  Indeed, this
is consistent with the fact that the Milky Way contains supergiants as late as M5,
while the latest-type supergiants known in the SMC 
(until now) were M2-3 (Humphreys 1979, Elias et al.\ 1985; Massey \& Olsen 2003; 
Levesque et al.\ 2006, hereafter Paper II).

Because of their extreme properties, the class of red supergiants includes a number of
oddballs.  In some cases, the peculiarities are due to a star's
binary nature, as in the case of VV Cep, whose spectrum 
shows emission lines of H and [Fe] in addition
to the  TiO bands that characterize an M star.  VV Cep is the archetype of
eclipsing systems consisting of
an M-type supergiant with a B-type companion. (For a nice review which has
stood the test of time, see Cowley 1969).  The star $\zeta$ Aur presents a similar, but 
less extreme, example (Cowley 1969, Wright 1970).  In other cases, the oddities (such as IR excess, photometric
variability and/or low-excitation emission lines) are
linked to extreme circumstellar environments, as is the case for VY CMa  and NML Cyg
(Hyland et al.\ 1969).  The circumstellar material associated with these stars
is indicative of extraordinary mass loss, due to some not yet understood instability.

These exceptional stars
are important not only for their own interesting astrophysics, but for what
 they can tell us about red supergiants in general.  For instance,  in VV Cep and $\zeta$ Aur, 
 the B star can be used as a probe of the atmospheric and wind structure in RSGs
(see, for example, Harper et al.\ 2005, Hack et al.\ 1992, and Schroeder 1985).
Massey et al.\ (2006) argue that VY CMa and NML Cyg may be
examples of normal RSGs caught during an unusually unstable period during which
the stars are losing copious amounts of mass, and their study thus sheds light on
an important, but short-lived, phase.

In this paper, we call attention to an equally unusual RSG in the SMC,   
HV~11423 
($\alpha_{\rm J2000}$=01$^h$00$^m$55.16$^s$,
$\delta_{\rm J2000}$=-71$^\circ$37'53.0"  from the UCAC2, Zacharias et al.\ 2004).
We have found that this star has changed its spectral type from K0-1~I to M4~I
and back again during the past two years, and that in late 2001 it was M4.5-5~I.
The M4 and M4.5-5~I types are considerably later than
the M2~I type cited as the latest spectral subtype found for SMC RSGs
by Humphreys (1979) and Elias et al.\ (1985);
Massey \& Olsen (2003) found only one star that might be M3~I.  
We will discuss other late-type SMC and LMC RSGs elsewhere (Levesque et al.\ 2007).
The temperature
corresponding to an M4-5~I type is  considerably cooler than what evolutionary
theory predicts for this metallicity (see Paper~II and discussion below), and
places the star well beyond the Hayashi limit, where stars cannot be in
hydrostatic equilibrium.
Furthermore, although 
{\it photometric} variability at the $\sim1$~mag level (in $V$) is fairly common for RSGs, 
{\it spectral} variability even by as much as a single spectral subtype (i.e., M1.5~I to
M2~I) is not. For instance, although Betelgeuse is well-known to vary
photometrically by as much as a magnitude, Goldberg (1984) failed to find any
sign of spectral variability; the star has remained an M2~Iab throughout many
decades of observations.  Even the highly unusual star VY CMa, which shows 3.5~mag
changes at $V$,  has remained constant in spectral type during 40 years of
intensive spectroscopic monitoring (Wallerstein 1958,
1977; Wallerstein \& Gonzalez 2001).  A change from K0-1~I to M4-5~I is without
precedent for a supergiant.

The star was listed as the brightest red star in the SMC by van
den Bergh (1968) and by Sandage
\& Tammann (1974), although subsequent work has revealed a few more SMC RSGs
that are equally bright, such as [M2002] SMC 06988 (see Massey \& Olsen 2003).
Sandage \& Tammann (1974) suggested using such stars as distance indicators.
The variability and ``very red" color was first described by Nail (1942) and
Shapley \& Nail (1951).  
The star is cataloged as [M2002] SMC~050028 in Massey (2002).  

\section{Observations: Old and New}

\subsection{Spectral and Photometric Variability}

In the course of an observational program to determine the physical properties
of RSGs in the Milky Way (Levesque et al.\ 2005, hereafter Paper~I) and 
Magellanic Clouds (Paper II) we chanced upon this 
remarkable star.  It was originally included in our SMC observing list as it had
been classified as an M0 supergiant by Humphreys (1979) and Elias et al.\ (1985) based
upon (photographic) spectra taken in 1978 and 1979 using the RC Spectrograph
on the CTIO 4-m.  (A note in their table adds that there was a ``warm continuum
at the blue end of spectrum".) They had observed the star
either because it was a Harvard variable, or from a preliminary list of RSG candidates
in the SMC that was eventually published as Sanduleak (1989), where
the star is designated as ``SkKM~205".  The photoelectric photometry reported by
Humphreys (1979) and Elias et al.\ (1985) shows that the star varied from $V=12.52$
(Oct 1978) to $V=11.77$ (Oct 1979) 
while staying essentially constant in $B-V$ (1.82-1.84). Although it is not clear from
these papers, examination of the CTIO observing schedules for the period in question
suggest that the spectra were obtained within a few days of the reported photometry.

Our first spectra of this star were
taken on 2004 December 2 (blue) and 2004 December 4 (red) with 
the CTIO 4-m and RC Spectrograph using the setups described in
Table~\ref{tab:setups}; further details are given in Paper~II. 
 We were quite surprised to find that the star was of early
K-type.  A comparison with spectral standards described in Paper~I suggest a
spectral type of K0-1~I.   Upon noting this discrepancy with the Humphreys (1979)
M0~I type, we were concerned that we might have observed the wrong star.  However,
we were forced to reject this possibility for the following reasons. First, the blue and
red spectra are clearly of the same object despite having been observed on different
nights; the features in the region of overlap are the same depths, and the flux levels
of the calibrated spectra agree to better than 2\%, so if the wrong object was
observed, it was observed on both nights.  That is unlikely, but would be possible
if (for instance) there was a mistake in the coordinate file used for pointing the telescope.
However,
we can eliminate that possibility as the coordinates in image headers, 
which reflect where the telescope was actually
pointed when the observations were made, agree with the 
correct coordinates to 3-5", as they do for the objects observed
directly before and afterwards.  Furthermore,
the fluxes we measure from our spectra correspond
to a $V$ magnitude of 11.84, consistent with the star's photometry.
The color we derive from our spectrophotometry corresponds to $B-V=1.85$, in
agreement with the literature values for this object.
Examination of the catalog of SMC stars by Massey (2002) shows
that HV~11423 is the only object this bright within 8'. Taken together, there seems to
be no question that the object we observed was the one we intended.

\begin{figure}
\epsscale{0.9}
\plotone{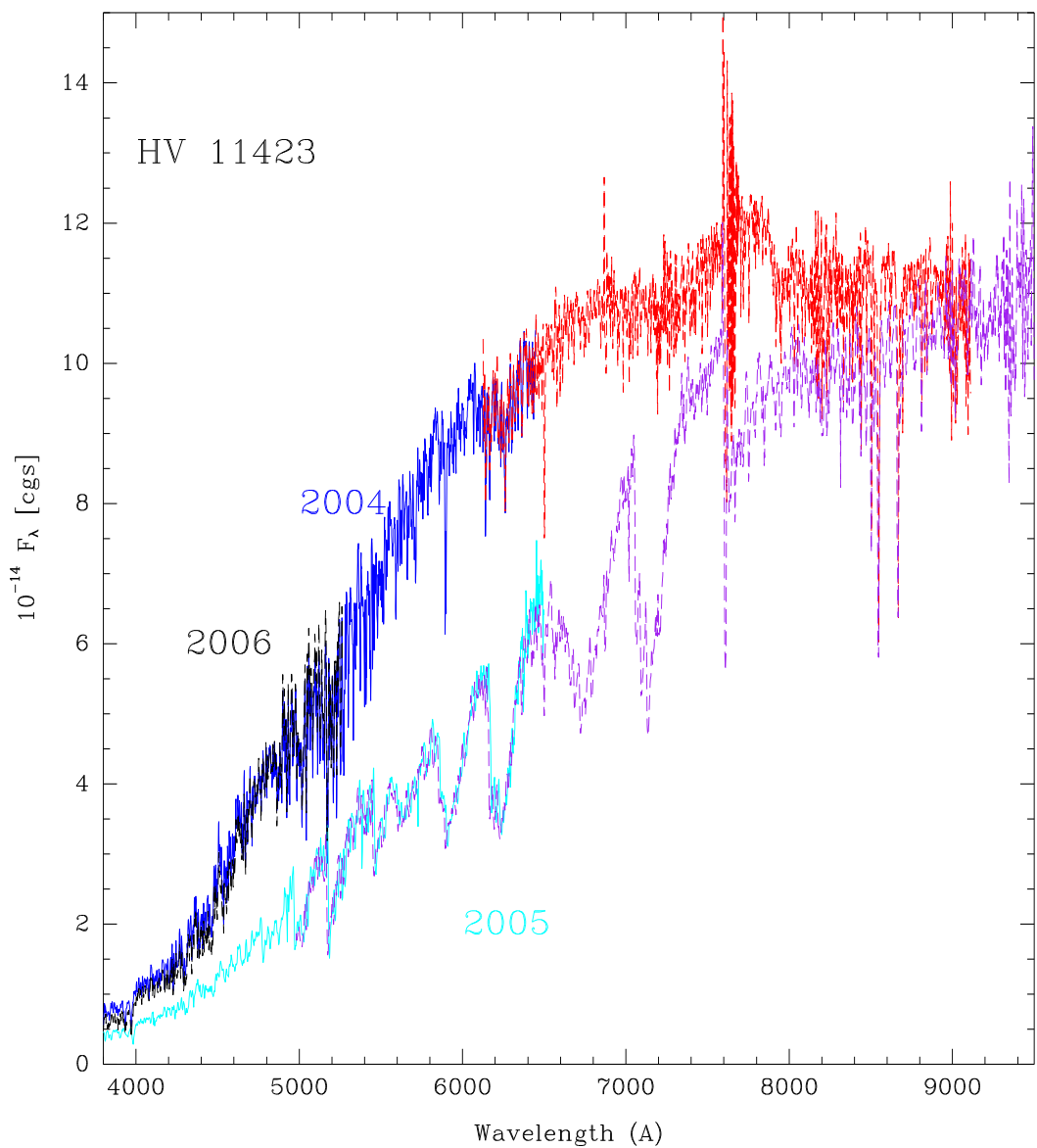}
\caption{\label{fig:spec} Spectra of HV 11423.  The upper
spectral energy distribution is from December 2004, with color used to 
distinguish the blue and red spectra taken a few days apart.  The spectral type
is K0-1~I.  Superposed in black
 is the 1.5-m SMARTS spectrum obtained in September 2006, which matches the 2004 observation
 very well. The lower spectral energy distribution is from December 2005, where
again color is used to show the individual (blue and red) observations.  The spectral
type here is M4~I, as evidenced by the very strong TiO molecular bands.   No 
adjustment in absolute level has been made for any of the four observations, showing
that the star was significantly brighter during the 2004 observation in the blue and 
visible, but that the star was of comparable brightness in the far red. The strong feature at
$\lambda 7600$ is the telluric A band. }
\end{figure}

We decided to re-observe the object the following year, and obtained spectra of
similar quality (cf., Table~\ref{tab:setups})
on 2005 Dec 20 (red) and 2005 Dec 21 (blue).  
If we were surprised by the K0-1~I spectrum
of 2004, imagine our reaction to the 2005 spectrum, where the
two are compared  in Fig.~\ref{fig:spec}.
The star is of spectral type M4~I,  the latest spectral type seen
for a supergiant in the SMC (Paper~II; see also Elias et al. 1985).  
The spectrophotometry indicates the star was now fainter
($V=12.46$) but with a  similar $B-V$, about 1.93.  This value is again similar to photometry
carried out at the same time, as described below.
Again the fluxes of the blue and red exposures agree
extremely well (as is obvious from the figure), and the coordinates in the headers
agree with each other and the target position to within 3-5", as do those of the
objects observed before and afterwards.  We conclude that the spectral type
changed from K0-1~I to M4~I between Dec 2004 and Dec 2005.  We note that the
star's flux remained unchanged in the far red (i.e., 9000\AA).

\begin{figure}
\epsscale{0.8}
\plotone{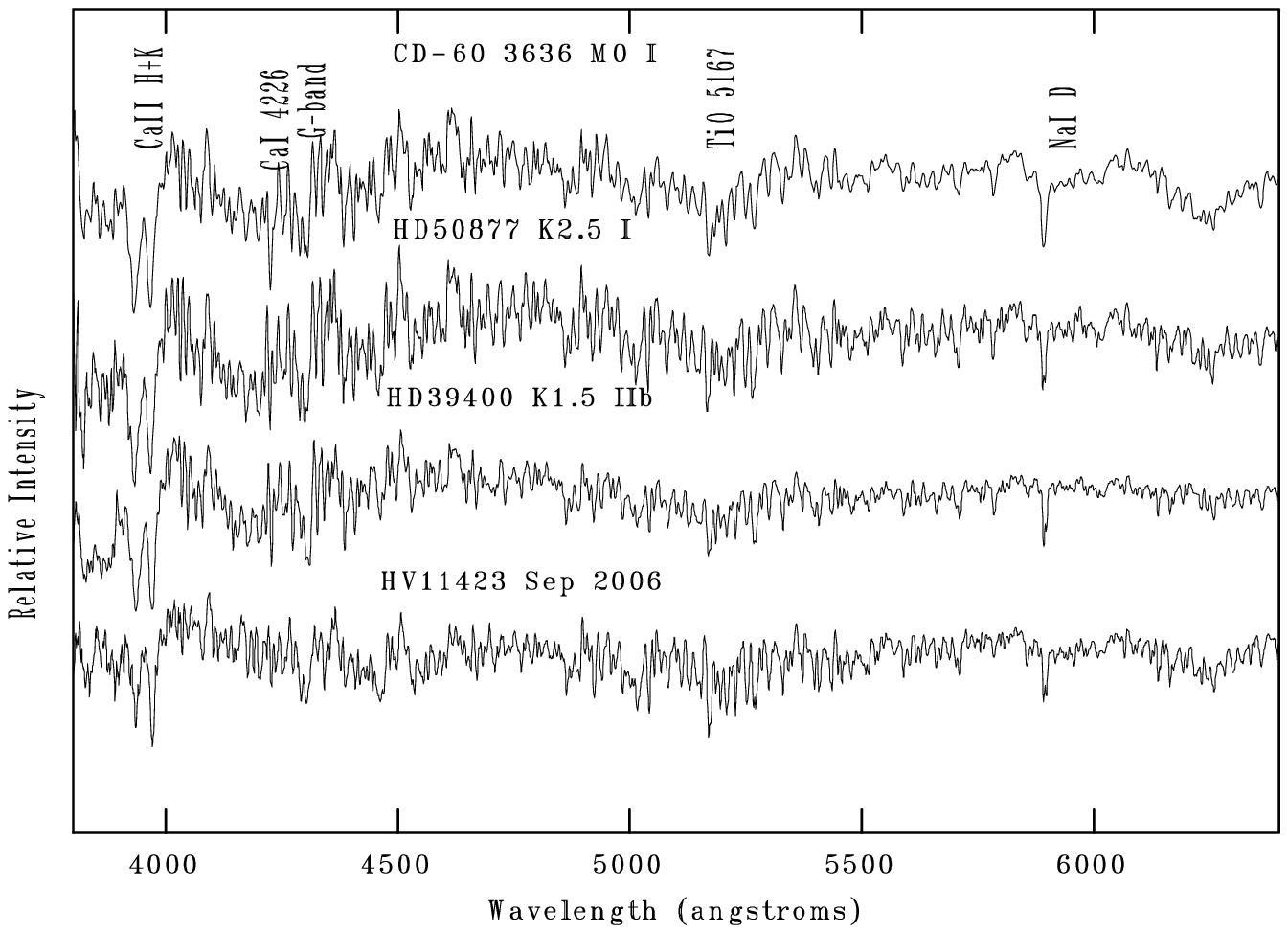}
\plotone{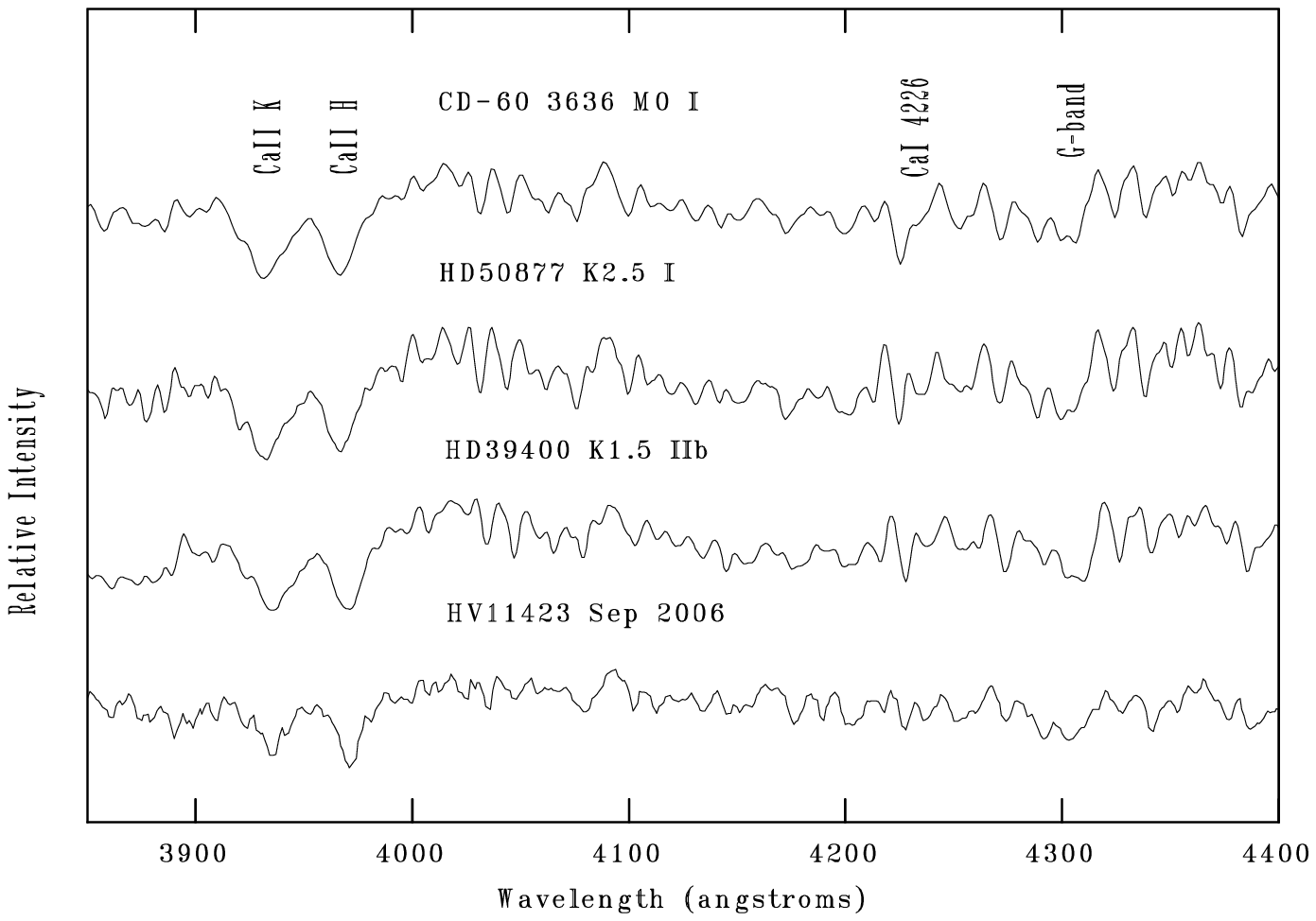}
\caption{\label{fig:kstar} Spectrum of HV 11423 in its K0~I state compared to
three spectral standards.  The lack of TiO $\lambda 5167$ in the top figure argues
that the September 2006 (and December 2004) spectra were considerably earlier
than the 1978 and 1979 M0~I spectral type given by Humphreys (1979) and
Elias et al.\ (1985).  The weakness of the CaII H and K lines, weakness of CaI $\lambda 4226$, and the overall appearance of the G band (lower figure) argues that the
star is K0~I.  For standards, we use spectra obtained by Levesque et al.\ (2005, 2006),
of the stars HD 39400 (K1~IIb, according to Morgan \& Keenan 1973),
HD 50877 (K2.5~I, according to Morgan \& Keenan 1973, and Levesque et al.\ 2005),
and CD-60$^\circ$3636 (M0~I, according to Humphreys 1978).
}
\end{figure}

As another check, we obtained two spectra in late September 2006.
The first of these was obtained on the CTIO 1.5-m telescope on 25 September 2006
through the courtesy
of F. Walters and the SMARTS observing queue.  The observation covered
only the blue (details are given in Table~\ref{tab:setups}), and was
flux calibrated by the observation of a single spectrophotometric standard.  
Both the spectral features and absolute flux agree very
well with our 2004 observation, as shown in Fig.~\ref{fig:spec}.  The second
spectrum was obtained on 29 September 2006 
during engineering time on the fiber positioner
Hydra on the CTIO 4-m telescope (Table~\ref{tab:setups}); we are grateful to A. Walker for his
permission and encouragement to do so.  Due to limited time we were unable to
obtain a satisfactory observation of a flux standard, but the excellent agreement
in the spectral features between the two late September observations is evident
(Fig.~\ref{fig:kstar}) in the region of overlap.
We identify some of the stronger features in this figure to emphasize
the certainty of the early K-type  classification (December 2004 and September 2006).
The lack of TiO $\lambda 5167$ requires that the spectral type be much earlier
than the 1978 and 1979 M0~I type, while the weakness of Ca~II H and K
absorption, the weakness of Ca~I $\lambda 4226$, and the overall appearance
of the G band argues for a very early K type.  The lack of Balmer absorption lines, though, argues against it being a late G-type star.  See Morgan et al.\ (1943),
and Jaschek \& Jaschek (1990).

\begin{figure}
\epsscale{0.8}
\plotone{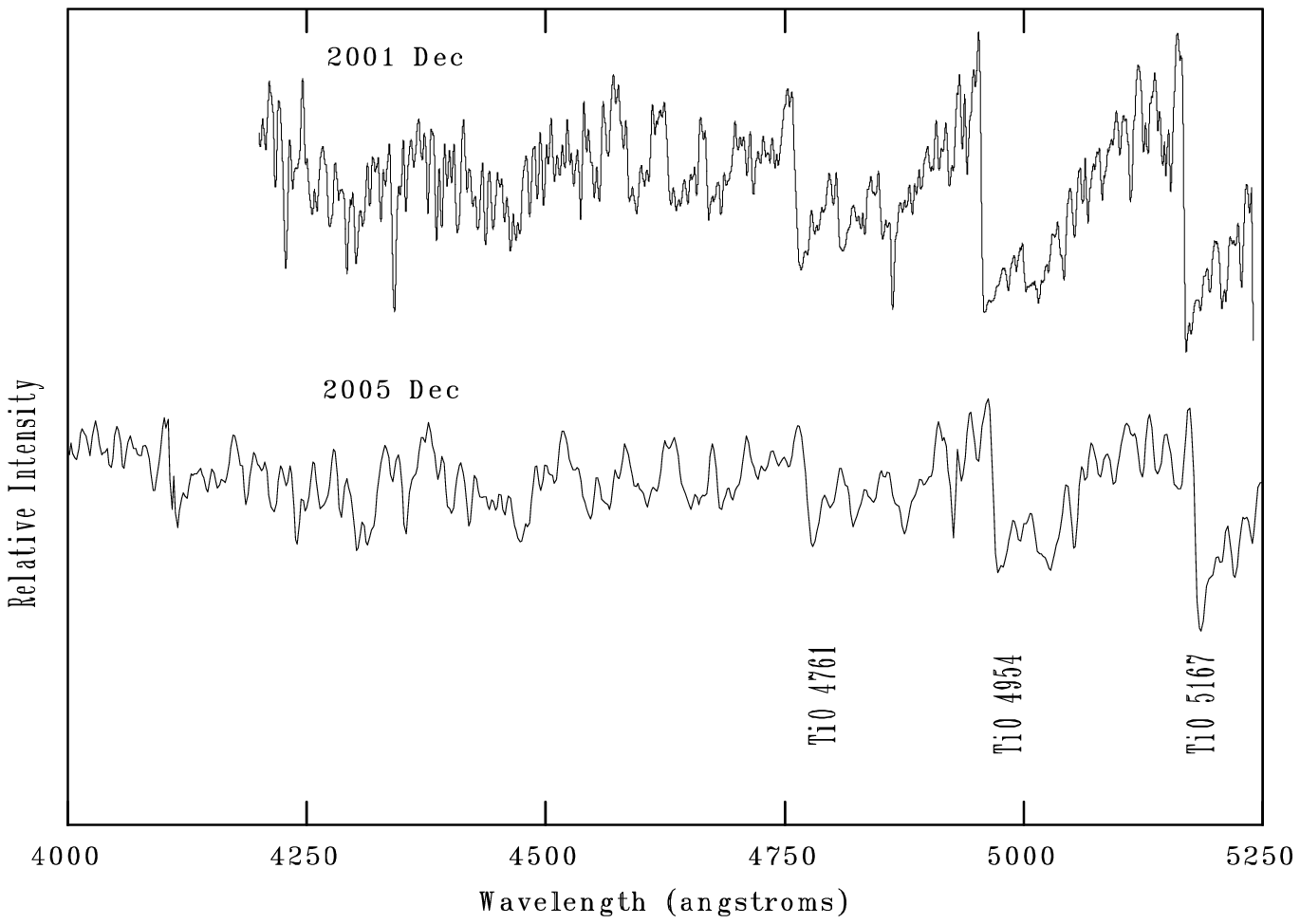}
\plotone{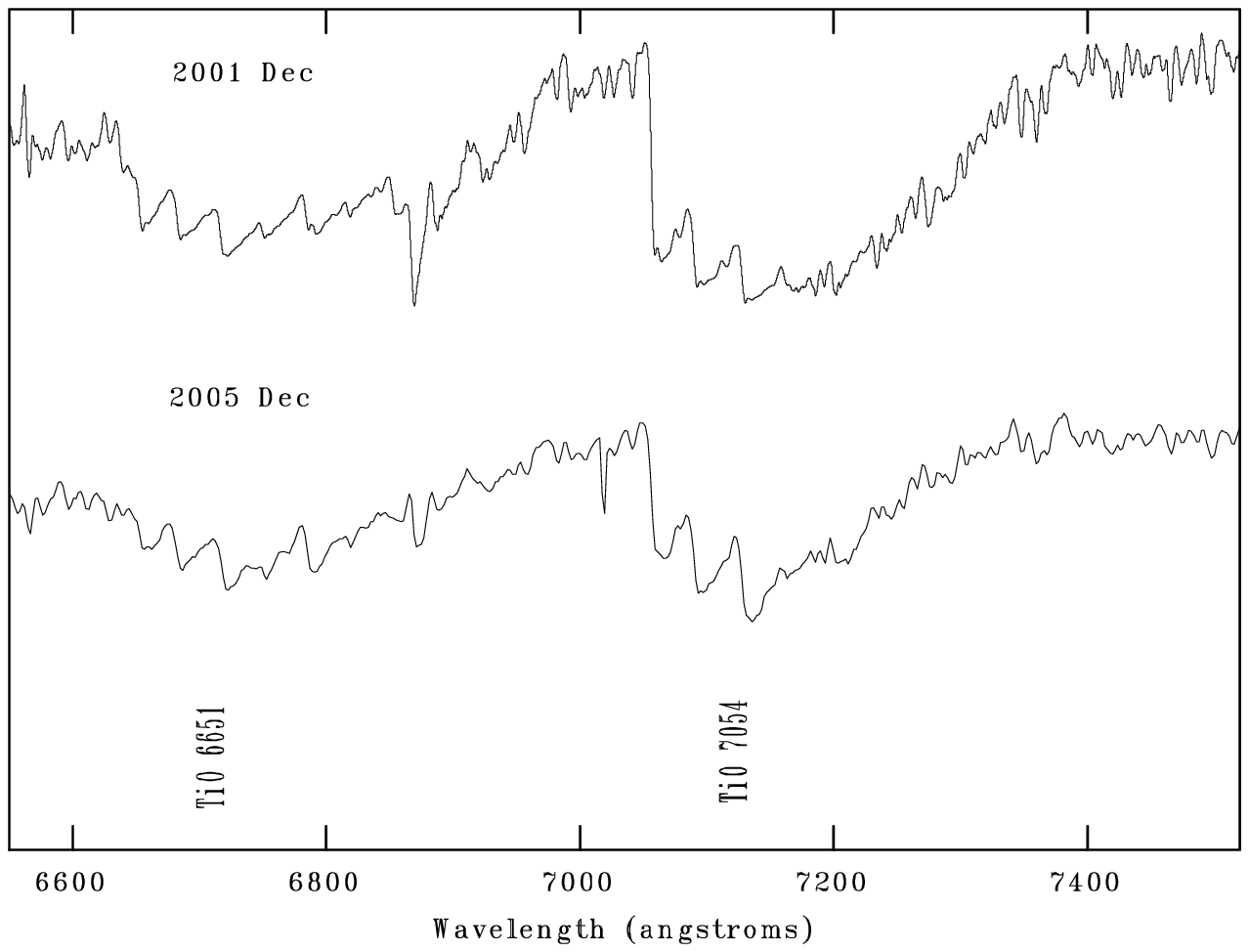}
\caption{\label{fig:ESO}Comparison of HV 11423 in its late-type state.  The
2001 December UVES spectra are compared to our
2005 December spectra.  The UVES spectra have been heavily smoothed to
approximately match our spectral resolution.  Both sets of observations have
been normalized for this comparison.  The TiO depths in the 2001 spectrum are
even deeper than in our 2005 (M4~I) spectrum, suggesting a type of M4.5-5~I.
}
\end{figure}

We were of course curious if we could find any additional spectral observations.
The ESO archive contained an echelle spectrum obtained with the VLT and
UVES on 5 December 2001, obtained under program ID 68.B-0530(B).   
The spectrum clearly shows that the star is
even later than the M4~I December 2004 type, probably M4.5-5~I.
We compare a small piece of the blue with our 2005 December
observation in Fig.~\ref{fig:ESO}.
As we shall shortly show, this spectrum happened to be obtained when the
star was at its faintest known $V$\footnote{Unfortunately the VLT observers did not
obtain any spectrophotometric standards during their run, and the setup was
not one of the standard ones for which the relative flux response is known.}.

\begin{figure}
\epsscale{1.0}
\plotone{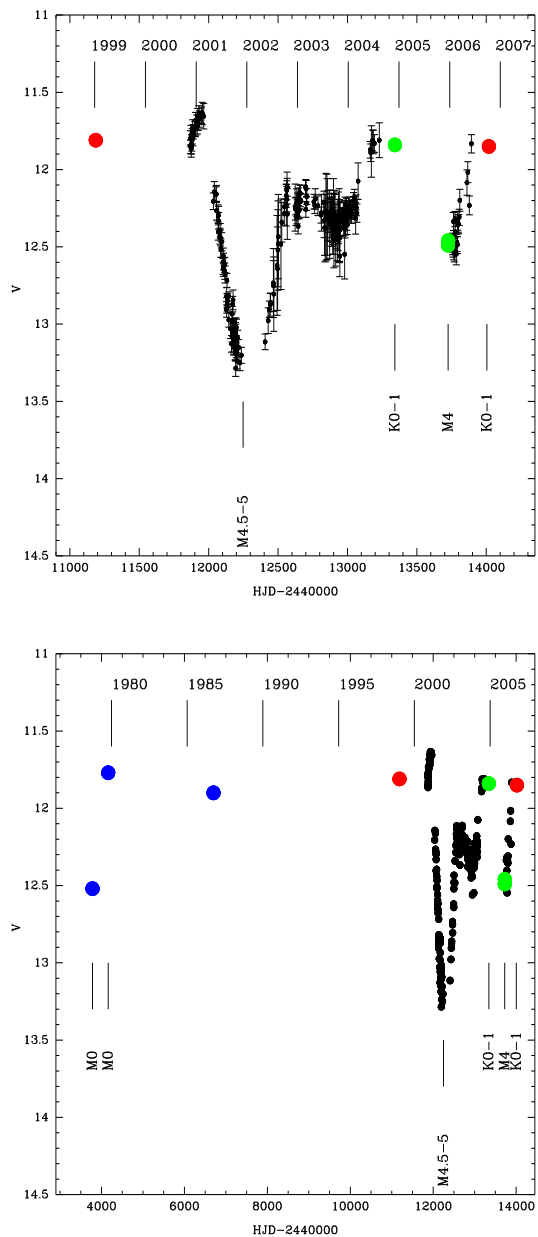}
\vskip -100pt
\caption{\label{fig:phot} Photometry of HV 11423.  The figure at the top shows
the ASAS data (error bars), the results of our spectrophotometry (green dots),
and two CCD $V$ values (red dots), from Massey (2002) and a measurement
new to this paper.
The figure on the bottom includes all of
these data (with the ASAS data now shown as small black points), and also three
photoelectric  $V$ band measurements (blue dots)
from 1978 (Humphreys 1979),
1979 (Elias et al.\ 1985), and 1986 (Maurice et al.\ 1989).
}
\end{figure}

In Fig.~\ref{fig:phot} {\it (top)} we show the star's $V$ band photometry (displayed
as error bars), where the data come primarily  from the All Sky
Automated Survey (ASAS) of Pojmanski (2002).   We also include the 
values derived from our spectrophotometry, which are
shown by green colored dots; these  are in good agreement with the ASAS values.  
The two red dots denote CCD observations: the $V=11.81$ value from
Massey (2002), obtained on 1999 January 8, and a newly obtained observation
on 2006 October 10 via the SMARTS 1.3-m queue, where we measure
$V=11.85$. (We are grateful to Charles Bailyn for arranging for this observation,
and to Suzanne Tourtellotte for processing the images through the Yale pipeline.)
In Fig.~\ref{fig:phot} {\it (bottom)} we include the photoelectric photometry
from 1978 and 1979 values 
($V=12.52$ and 11.77  from
Humphreys 1979 and Elias et al.\ 1985, respectively), as well as a 1986 photoelectric
measurement ($V=11.90$) from
Maurice et al.\ (1989), with all three  shown in blue.

It is clear from this that the star varies nearly 2 mag in $V$.
The star had a deep minimum in December 2001 ($V=13.5$), and
a ``mini-minimum" in October 2003.  The star was observed coming out of
a minimum in December 2005.
If that was a deep minimum similar to that seen in December 2001, then
folding the light curve based upon these
two minima suggest a 1295 day (3.5 yrs) period.  However, this is at
variance with the 1999 CCD observation, which would be nearly 0.5 mag too
bright for that period.   Alternatively, if the minima were very unequal in depth,
and the ``mini-minimum" correspond to the same phase as the deep minimum,
then there might be a 700 day period.  The rising part of the
light curve observed in September 2006 might also match this, but clearly several
more years of photometric monitoring will be needed before we can determine
if there is any periodicity or not.
We can say
with certainty that the two K0-1~I spectra were both at times when the star
was at its brightest, but that the M4~I spectra were taken when the star was fainter:
the 2001 spectrum was taken at nearly minimum light.  The two M0~I spectra in 1978 and 1979 were both
obtained when the star was bright (1979) and relatively faint (1978), if indeed
the published photometric and spectra data were contemporary.

Various $K$ band measurements exist in the literature, dating back to
mid 1977 (Glass 1979) and continuing to 1998 with the 
DENIS (Cioni et al.\ 2000) and 2MASS (Skrutskie et al.\ 2006) surveys.  
During that time the star stayed essentially constant at $K$, with values spanning
the range 7.86-8.08, when transformed to the standard system of Bessel \& Brett (1988)
using Carpenter (2001).  This consistency is in accordance with what we see in our
spectral energy distributions (Fig.~\ref{fig:spec}),  where the fluxes agree
in the far red despite large differences at shorter wavelengths.  
It is also consistent with the general 
supposed trend that RSGs show considerably smaller 
variations at $K$ than at $V$
(Josselin et al.\ 2000), although there are clearly exceptions, such
as VY CMa (Massey et al.\ 2006).
We adopt the average,
$K=7.95$.

The star's radial velocity was measured in August 1983 as
107.7$\pm1.1$ km s$^{-1}$ by Maurice et al.\ (1987) using the 
CORAVEL photoelectric scanner.  The radial velocity of this
star was considered by them to be a
``high residual", and indeed the velocity is low compared to the 158 km s$^{-1}$
of the center of the SMC, and the distribution of radial velocities of other SMCs (see 
Fig.~1 in Massey \& Olsen 2003), although above the 100 km s$^{-1}$ value used
by Massey \& Olsen (2003) to
assign SMC membership.  We have two measurements, 
based upon cross-correlating the region around the Ca II 
triplet with SMC/LMC RSGs from
Massey \& Olsen (2003).  
The 2004 December value is $110\pm5$ km s$^{-1}$, in good agreement with the
CORAVEL value.   Our 2005 December value is $122\pm 10$ km s$^{-1}$, where the large
error is a reflection of the lower spectral dispersion.  We cannot rule out velocity variations
of the star, but  our data do not show any evidence of them.

Observationally, HV~11423 is remarkable in one other way: as noted
by Loup et al.\ (1997), it is an IRAS source (F00592-7153), with
an (uncorrected) flux at 12$\mu$m of 0.114 Jy (Moshir et al.\ 1990).
Of the 163 stars listed by Massey \& Olsen (2003) as potential or
confirmed RSGs in the SMC, only 3 others are IRAS sources.  We list
these in Table~\ref{tab:iras}.  Unfortunately, there is no
information about the variability of the star in the mid-IR; the
object was about half a degree outside of the Midcourse Space
Experiment {\it MSX} survey region (Price et al.\ 2001, Egan et al.\ 2003), and
is also outside of the SMC {\it Spitzer} survey.  Nor are there any mid-IR imaging
of the object in the Gemini Science Archive. 
Observations with {\it Spitzer}
would be highly desirable, and we plan to propose these.

\subsection{Summary of Observations}

We briefly summarize the observational facts here.  The star
has been well observed photometrically in $V$ since late 2000, and is found
to vary  from 11.5 to 13.4.  There was a deep minimum occurring in 
December 2001 (Fig.~\ref{fig:phot}) and a spectrum taken at that time at the VLT 
shows
a very late-type spectrum, M4.5-5~I.  In November 2004 the star was at
its brightest, and the spectrum was a K0-1~I.  A year later, in December 2005,
the star was faint but brightening 
and the spectrum was observed to be M4~I.  In September
2006 the star was again bright, with a spectrum of K0-1~I.  The simple picture that
the star is an M4-5~I when faint, and a K0-1~I when bright, is at variance
with the 1978 and 1979 observations by Humphreys (1979) and Elias et al.\ (1985).
They identified the star as spectral type M0~I both times, despite the fact that
the star had $V\sim 12.5$ for the 1978 observation and $V\sim 11.8$ for the 1979
observation.

In contrast with the several hundred $V$ band measurements, there are only
a few $B-V$ measurements available, and these have remained relatively constant
at $B-V=1.82-1.93$, despite changes in the star's $V$ photometry and spectral type.
(We should note, however, that none of the $B-V$ measurements were at the
deep $V$-band minimum.)  There are only a few $K$ band measurements of the
star, but these have also stayed constant, to within uncertainties.  The star has a
radial velocity consistent with membership in the SMC, and is one of just four SMC
RSGs that were detected at 12$\mu$ by IRAS.

 \section{Physical Properties}

Following Papers I and II, we can estimate the physical properties in two ways,
using the spectrophotometry and the $V-K$ colors.  For  both we rely on the 
MARCS stellar atmosphere models (Gustafsson et al.\ 1975, 2003;
Plez et al.\ 1992; Plez 2003), where a grid of models have been computed
for $Z/Z_\odot$=0.2 (appropriate to the SMC; see Westerlund [1997]) for
3000-4500 K and surface gravities from $\log g=-1$ to +1 [cgs units].
We adopt a true distance modulus of 18.9, following Westerlund (1997) and
van den Bergh (2000).

\begin{figure}
\epsscale{1.0}
\plotone{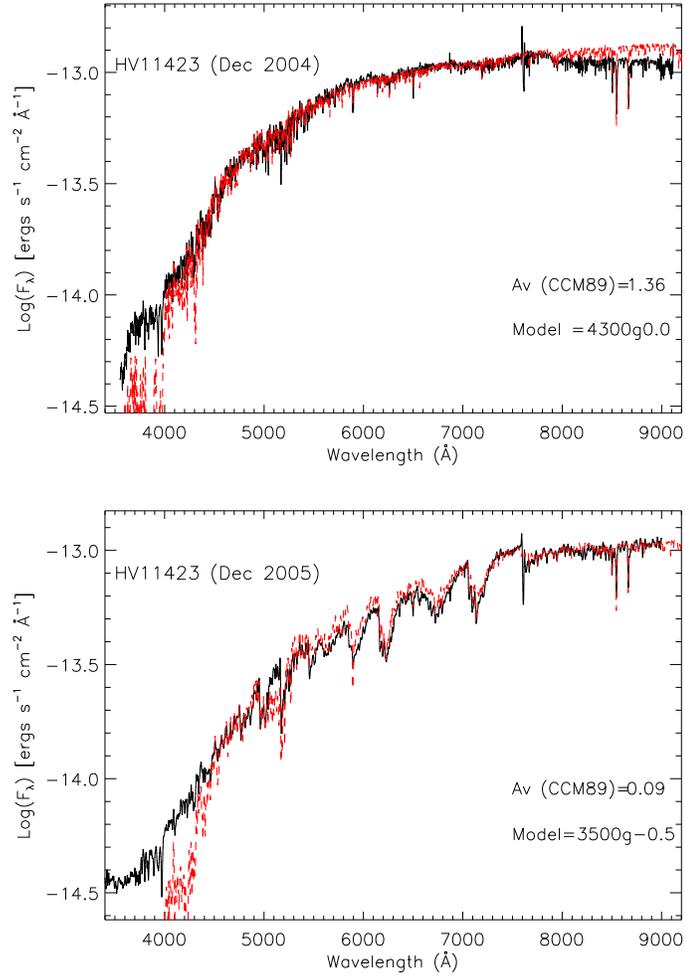}
\vskip -100pt
\caption{\label{fig:fits}Model fits to the spectra of HV 11423.  In the upper figure we show
the spectrum of the star (black) and the best fit model (red) for the December 2004 observation.
In the lower figure we show the same for the December 2005 observation. }
\end{figure}

First, we fit the synthetic spectra to the observed spectral energy distribution,
beginning with the $\log g=0.0$ models and allowing the effective temperature
of the model and the amount of extinction, $A_V$, to vary until we obtain a
satisfactory fit.  For this, we adopt the standard
Cardelli et al.\ (1989) reddening law.  For an early K-type star, the
procedure is not very exact, with an uncertainty of 100 K in the derived effective
temperature, owing to the lack of strong, temperature-dependent spectral features
at this dispersion, and the overall degeneracy between reddening and effective temperature
in terms of the continuum shape.  For late-type M supergiants, the precision is considerably
better, about 25~K, owing to the sensitivity of the strong TiO bands to effective temperature.
Once a satisfactory fit is obtained, we compute the absolute visual luminosity $M_V=V-A_V-18.9$, and then determine the
bolometric luminosity using a relationship
between the bolometric correction and effective temperature from the MARCS models (Paper~II).
We can then use an
approximate relation between mass and luminosity (see Papers I and II), and the star's location
on the HRD, to estimate the expected value for $\log g$.  If the value is significantly different (0.3~dex)
from 0.0, we repeat the fit using models with more appropriate surface gravities.  In practice, changing
$\log g$ has a minimal effect on the derived $A_V$, and no effect on the deduced $T_{\rm eff}$.
The spectral fits are shown in Fig.~\ref{fig:fits}, and the results given in Table~\ref{tab:results}.

We note that neither spectral fit is very satisfactory, and neither
is as good as what we usually achieve (Papers I and II).  
The fit to the 2004 spectrum (K0-1~I)is shown in the upper panel.  The extinction used
is the best compromise. A lower value ($A_V=1.25$) would provide
better agreement in the far red ($>$8000\AA),  but would result in a much 
worse fit in the 6000-7000\AA\  region.
Changing the temperature to either higher or lower values results in a poorer fit, 
but a change by 100 K in either direction is
only marginally worse.  The fit to the 
2005 spectrum (M4~I) shown in the lower panel does better in terms of fitting both the continuum
and TiO line depths, and the latter constrain  $T_{\rm eff}$ to $\pm$25~K of the 3500~K value.  The slight mismatch at 5200\AA\ is an artifact of the smoothing process
of the synthetic spectra; see discussion in Section 3.2 of Paper II.
In neither case is the extinction value characteristic of the massive stars in the SMC, where the
distribution is tightly peaked at $A_V=0.28$ (Massey et al.\ 1995).  In both fits, there is significantly
more flux from the star in the near-UV than in the models.  This problem, though, occurs for most
of our RSGs fits, and Massey et al.\ (2005) argue that this is due to the scattering of light by circumstellar
dust.  

The second method uses the $V-K$ photometry to obtain an effective temperature.  We adopt
$A_V$ from the spectral fits, and determine $(V-K)_0=(V-K)-0.88A_V$, following Schlegel et al.\ (1998).
We next determine $T_{\rm eff}$ using the calibration we determined in Paper~II between the
effective temperatures of the
MARCS models and their $(V-K)_0$ colors.  We then determine the star's absolute K-band
luminosity $M_K=K-0.12A_V-18.9$, and apply a K-band bolometric correction (which will
be positive) to determine the bolometric luminosity.  

We give the results of our fits in Table~\ref{tab:results}.  For each year, the agreement between
the spectral fit and the results from $V-K$ photometry is remarkably good.  
In Paper~II we found
that there tended to be a systematic difference between the $T_{\rm eff}$ values determined
from spectral fitting and from the broad-band $V-K$ photometry, with the latter 
105 K higher (on average)
for SMC stars.  This is  similar to the 180 K (2004) and 135 K (2005) differences seen here.  Despite
this, the bolometric luminosity of the star are nearly identical for the two methods.

The good agreement in effective temperatures between spectral fittings and $(V-K)$ for the 2004 K0-1~I observation
{\it requires} the unusually high reddening we find from the model fitting: if the reddening
was more similar to that of the average SMC ($A_V=0.3$) rather than $A_V=1.4$, the 
effective temperature from $V-K$ would be about 500 K cooler, and it would disagree considerably
with the spectral fitting.    So, we feel that the high reddening for the 2004 K0~I observation
must be real. The reddening for the 2005 M0~I observation is slightly
spuriously low: if we adopted a slightly higher surface gravity model ($\log g=0.0$)
our best fit then requires $A_V=0.31$.  This would raise the $V-K$ temperature only slightly,
by 80 K.

More remarkable is the fact that despite the tremendously large change in apparent spectral type
(K0-1~I to M4~I) and the 1.9~mag change in $V$, the derived bolometric luminosity is essentially the same for the 2004 and 2005 
observations: the change in $V$ is largely
compensated by the change in $A_V$ and the change in the bolometric correction.

Let us briefly consider the 1978-1979 observations.   For $V=12.52$ (October
1978), the implied temperature from $V-K$ is 3680 K, where we have assumed that the star's reddening
is average for the SMC.  Given that, we find it surprising that the 1978 spectrum was that of an M0~I star.
That temperature would correspond to that of an M2~I, and if we correct by 105~K (as per above), more
like an M3.5~I.  The 1979 observation, however, in which $V=11.77$, yields an effective temperature
of 3995~K, assuming $A_V=0.3$,  more like an early K-type (see Paper~II), rather than the observed M0~I type.
We conclude that if the published spectra and photometry were obtained 
contemporaneously, as Elias et al.\ (1985) imply, then the visual extinction
must have  changed
considerably between the 1978 and 1979 observations. Still, that appears
to be at variance with the constant $B-V$ observations.

Finally, we note that at the time of the 2001 December M4.5-5~I observation of the star
would have been made when $V\sim 13.4$ (Fig.~\ref{fig:phot}, implying
$V-K\sim 5.5$.
 If we deredden this by $A_V=0.3$ as above, then the effective temperature
 would have been 3420~K.  Corrected by 105 K, would lead to a temperature of
 3315 K, making this one of the coolest RSGs known in any galaxy (see Table 4
 of Paper II).  Were the
 reddening higher than the  $A_V=0.3$ we assumed, then the temperature would be
 warmer, but such a cool temperature is consistent with the
 very late (M4.5-5~I) spectral type.

\section{So What's Going On?}

\begin{figure}
\epsscale{1.0}
\plotone{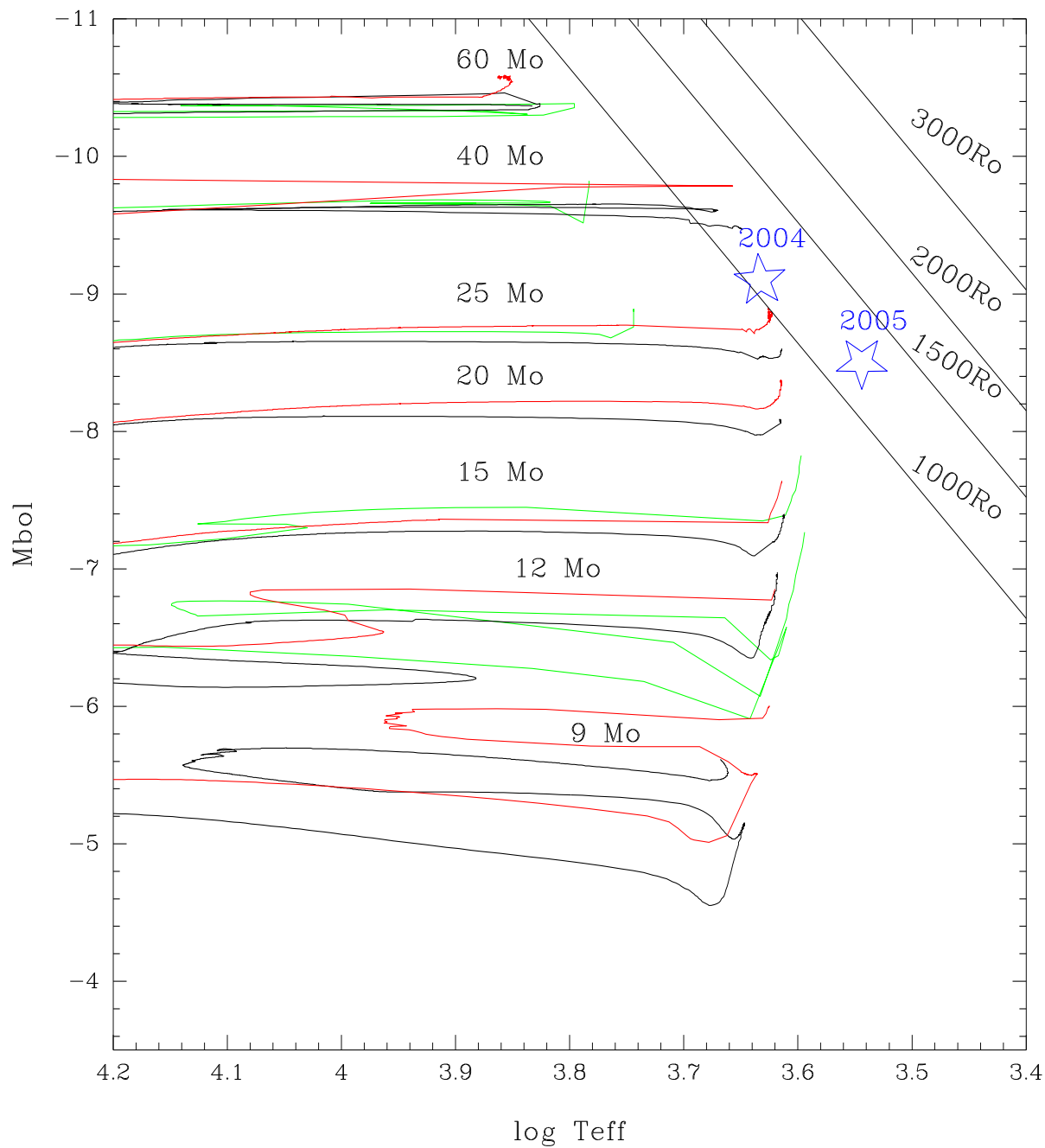}
\caption{\label{fig:hvhrd}The location of HV~11423 in the HRD.  The two purple stars denote the 
2004 (warmer, more luminous) and 2005 (cooler, less luminous) physical properties of
HV~11423.  The Geneva evolutionary tracks are shown, labeled with the (initial)
masses.
The older, non-rotation evolutionary tracks of Chabonnel et al.\ (1993) are shown in green.
The newer tracks (when available) from Maeder \& Meynet (2001) are shown in
black (zero rotation) and red (300 km s$^{-1}$ initial rotation).}
\end{figure}

The 2001 and 2005 M4-5~I spectral types are the latest seen in the SMC,
and the implied
cool effective temperature places the star clearly in the Hayashi ``forbidden zone", as shown in Fig.~\ref{fig:hvhrd}: the star is well to the right of where the evolutionary tracks end.
Such stars should not be in hydrostatic equilibrium, and in the case of HV~11423 we suspect
that this is just what is happening: that the star is undergoing huge bursts of mass-loss combined
with spectral (and photometric) variability.  Dust condenses a few stellar radii
away from the star, and as 
Massey et al.\ (2005) showed, this dust can substantially alter
the circumstellar environment, creating extra extinction and reddening.   Such dust would have little
effect on the $K$-band magnitude, but would be observable further in the infrared.
Consistent with this is the fact that HV~11423 is an IRAS source.

We can estimate the dust mass-loss rate from the ratio of the flux at 12$\mu$m (due to dust) to the flux
in the K-band (due to the star), following Josselin et al.\ (2000).  The 12$\mu$m magnitude [12] is
defined as $-2.5\log(S_{12}/28.3)$, where $S_{12}$ is the IRAS flux at 12$\mu$ (in Jy, and uncorrected
for color), and  the numerical factor has been chosen so that $K_0-$[12]=0 for
a 10,000 K star.  Stars with positive $K_0-$[12] will have some 12$\mu$m excess, as the index has little
sensitivity to effective temperature.  
Josselin et al.\ (2000) derive $$\log\dot{M_d}=0.57 (K_0 - {\rm [12]}) - 9.95,$$ where $\dot{M_d}$ is the dust
production rate in units of $M_\odot$ yr$^{-1}$.
For $S_{12}=0.11$~Jy, [12]=5.99.  We assume  $K_0=7.92$, based upon dereddening $K=7.95$ by
the average SMC $A_V=0.3$ ($A_K=0.12 A_V$), and thus $K_0-$[12]=1.93.  The implied 
dust production rate is $1.4 \times 10^{-9} M_\odot$ yr$^{-1}$.

This value is fairly high for a Galactic RSG, but not phenomenally so---there are RSGs with $\dot{M_d}$
4-5 times higher, and indeed we would expect a higher rate if this were a Galactic RSG with such a 
high bolometric luminosity.   But, in the SMC, where the metallicity is low, this rate is apparently as
high as seen in any RSG.  In Table~\ref{tab:iras} we list all of the SMC RSGs and RSG candidates
which are associated with IRAS sources at 12$\mu$m.  Given that $K$ is likely variable for all of these
sources at the 0.2~mag level, and that the IRAS fluxes are uncertain at a similar level, we conclude
that all four of these SMC RSGs had similar dust production rates at the time the IRAS data were
taken.   

Is this amount of mass-loss sufficient to account for the change in the observed
extinction between our 2004 and 2005 observations? ($\Delta A_V=1.25$~mag,
according to Table~\ref{tab:results}).
 Massey et al.\ (2005) note many examples where RSGs in Galactic
clusters and OB associations show extinction in excess of 1~mag compared to
early-type stars in the same region, and show that this extra extinction is correlated
with the bolometric luminosity and dust production rates.  Massey et al.\ (2005)
describe a very simple thin shell 
approximation that shows that the amount of extinction is of the right order of magnitude
using very conservative assumption.  We do not know the distance from the
star that the dust will condense, nor do we know the thickness of the shell
(corresponding to the duration of episodic mass-loss), but if we take reasonable
values (10 yrs and 3 stellar radii, see Danchi et al.\ 1994), we expect about 1.5~mag
of extinction at $V$ for a dust production rate of $1.4\times 10^{-9} M_\odot$ yr$^{-1}$.
Of course, this comes from a single observation, and
 without continued mid-IR observations, we do not know how 
representative this dust production rate is.

As described in some detail by Massey et al.\ (2005),
for grain sizes that 
are typical of those found in the ISM, the primary effect of such
a dust shell in the yellow-red will be to add extra reddening.  If the particle sizes
were larger, we would expect grey extinction, that would be harder to notice.
The variability in $V$ {\it might} be due to the variability in the visual extinction caused
by dust,
as is the case with R Coronae Borealis type stars (Hecht et al.\  1984),  and
some long period variables  (i.e., L$_{\rm 2}$ Pup, Bedding et al.\ 2002), as
the extra $A_V$ we found (Table~\ref{tab:results}) is of the correct order of magnitude
as the star's observed variability in $V$.
However, if the star's bolometric luminosity is essentially constant, as our findings
seem to indicate, despite the large swings in effective temperature, then we would 
expect differences in the bolometric correction to account for much of the
variability at $V$.  

The near constancy of $B-V$ despite the large changes in $V$ is
consistent with the large changes in the effective temperature being
the dominant source of $V$-band variability.  A change in effective
temperature from 4300~K to 3300~K (such as we measure from the
star's spectra) would change the intrinsic $B-V$ by only 0.07 mag,
according to the MARCS models.  In contrast, a change in $A_V$ of
2~mag (needed if variability in the dust was the culprit) would
change the star's $B-V$ color by 0.65 mag, if the ratio of total
to selective extinction were normal.  However, this might be
ameliorated by extra blue and near-UV light being scattered into
the line of sight by dust close to the star, as described by Massey
et al.\ (2005). Such UV excesses are known around R Coronae Borealis
type stars, and may explain why the (limited) $B-V$ data remains
relatively constant despite changes in the overall reddening in the
yellow-red part of the spectrum. 

We propose that HV 11423 is undergoing an unusual period of instability, 
during which its effective
temperature changes from 4300~K to 3300~K on a time-scale of a year or less.  
During these swings the star stays at constant bolometric
luminosity, or nearly so.  What causes the large variations in $V$?  Mostly this
is due to the change in effective temperatures, as is apparent from comparing the
physical properties associated with the
 2004 December and 2005 December observations (Table~2).  The star changed
 by only 0.6~mag in $M_{\rm bol}$ while staying at essentially the same radius.
 The absolute visual magnitude $M_V$ 
 (i.e., corrected for extinction) changed by 1.9~mag, due to the change in
 effective temperature.  The change in extinction by $-1.2$~mag helped ameliorate 
 the effect on $V$ in this particular instance, resulting in a $\Delta V$ of only
 0.6~mag.  But without the change in $A_V$ the variation in $V$ would have
 been nearly 2~mag. In general, when the star is
very cool we expect the star to be out of hydrostatic equilibrium
(as it is well into the Hayashi forbidden zone), and we can expect
that copious mass-loss, with associated dust production, occurs.  As Massey et al.\ (2005) showed,
such dust loss can lead to a magnitude or more of visual extinction. Neither the changing temperature,
nor the dust production, would have much affect at $K$, as the bolometric correction changes by
only 0.5~mag for this temperature range, and the extinction at $K$ is only 12\% of what it is at $V$.
The idea that the changing $V$ is {\it partially} due to dust extinction
may also help explain the complexity of the light curve.   It may also explain
how the star could be observed as M0~I both in 1978 and 1979 despite the
large change in $V$, although this seems to be in conflict with the published $B-V$
photometry.

We briefly considered an intriguing alternative, namely that the star is a
spectral composite binary.  If the system consisted of a late-type M star heated on
one side by a hot companion, this would provide an explanation for the changing
spectral type and $V$ brightness.
The apparent change in reddening might also be explained by this, 
as it would simply be an artifact of fitting the spectral energy distribution with
a model of a single temperature.  However, the explanation might
be hard to reconcile with the
lack of periodicity in the light-curve, and the intermediate
M0~I type that was observed both in 1978 and 1979.  It would also not explain
why the star is such a strong 12$\mu$m source.  Finally, it would mainly just add
a complication (binarity) to the principle mystery, namely how such
a late-type star could exist in the SMC.  An M4-5~I star in the SMC should be unstable,
and we consider it more likely that the variability is intrinsic to the cool temperature.
Still, radial velocity monitoring would be useful.

Spectral variability, such as what we have found here for HV~11423, has never
been associated with red supergiants, although it is reminiscent of the properties of variable stars found in other regions of the HRD.
For instance, the RV Tau variables are (low luminosity)
yellow supergiants which change their spectral types by
about one class, being F to G at minimum light and G to K at maximum.  These stars
also show large IR excesses due to dust (see, for example, Jura 1986).  Other
variables at the tip of the asymptotic giant branch, such as the Mira variables
also show spectral variability, as well as 
dust (see, for example, Whitelock et al.\ 2003).
The physics of these variables are not particularly well understood, and their
relevance to the spectral variability we see in this supergiant is not obvious.
However, Levesque et al.\ (2007) have recently studied a sample of other late-type
RSGs in the Magellanic Clouds.  At least two of these stars also show 
evidence of spectral variability on the time-scale of a few years, [M2002] SMC 046662
and LMC 170452.  

Stars that somehow find themselves in the Hayashi forbidden region of the H-R
diagram should struggle to quickly regain hydrostatic equilibrium, and that is consistent
with the fact that HV~11423 returns to an early K-type within months of being
in its late-type state in December 2005.  (We suspect that a similar thing happened in
December 2001.)  Why then was the star observed to be of intermediate type
(M0~I) both in October 1978 and October 1979?  Could this instability have just
recently set in?  If that is the case, then perhaps we are witnessing the early 
death throes of this RSG.  Kiss et al.\ (2006) have recently analyzed the historical
light curves of a large sample of Galactic RSGs, and they identify a small subset
that have large $V$-band variability, similar to what is seen here.  They propose
that this is an indicator of a ``super-wind" phase that precedes the supernova stage.
Clearly continued photometric and spectroscopic monitoring of HV~11423 is warranted.

\acknowledgements 

We gratefully acknowledge useful correspondence and conversations
about this interesting star with Drs.\ Beverly Smith,
Geoff Clayton, and Roberto Mendez   This work was supported by the NSF, through
grant AST-0604569.
This paper made use of data products from the Two Micron All Sky Survey, which is a joint project of the University of 
Massachusetts and the Infrared Processing and Analysis Center/California Institute of Technology, funded by the National 
Aeronautics and Space Administration and the National Science Foundation.
It also made use of observations made with the European Southern Observatory
telescopes  at Paranal Observatories under program ID 68.B-0530(B), and
obtained from the ESO/ST-ECF Science Archive Facility.
We also note that two of the critical follow-up observations were made as
``targets of opportunity" with the observing queues on the SMARTS 1.5-m and
1.3-m telescopes; the availability of this observing mode is of great benefits to such
projects.   We also greatly benefited from the All Sky Automated Survey of Pojmanski (2002),
a valuable resource.  We are grateful to Dr.\ Deidre Hunter for a critical reading of the
manuscript, and to Dr. David Silva for advice on reducing the archival ESO UVES 
spectrum.  An anonymous referee offered several useful remarks that helped us 
improve this paper.

\clearpage
\begin{deluxetable}{l c c c c c c c c}
\rotate
\tabletypesize{\scriptsize}
\tablewidth{0pc}
\tablenum{1}
\tablecolumns{9}
\tablecaption{\label{tab:setups}  Spectrograph Setups\tablenotemark{a}}
\tablehead{
\colhead{}
&\multicolumn{2}{c}{4-m Dec 2004}
&\colhead{}
&\multicolumn{2}{c}{4-m Dec 2005}
&\colhead{1.5-m Sep 2006}  
&\colhead{4-m Sep 2006}\\ \cline{2-3} \cline{5-6}
\colhead{Parameter}
&\colhead{BLUE}
&\colhead{RED}
&\colhead{}
&\colhead{BLUE}
&\colhead{RED}
&\colhead{}
&\colhead{}
&\colhead{}
}
\startdata
Grating/l mm$^{-1}$  &KPGL1(I)/632  &KPGLF(I)/632  & &KPGL2(I)/316 &KPGL2(I)/316 & 26/600 &KPGL3/316\\
Blocking Filter &GG345 &GG495 & &      BG39 &WG495 & None & None\\
Wavelength Coverage(\AA\ ) &3550-6420 &6130-9100 && 3400-6200 &5300-9000& 3540-5300 & 4200-6700 \\
Slit Width (''/$\mu$m) &2.5/375 &2.5/375 &  &1.5/225 &1.5/225 & 2.0 /110 &2.0 /300\tablenotemark{b}\\
Dispersion(\AA\ pixel$^{-1}$) &1.01 &1.04 & &1.99 & 2.00 & 1.48 &1.39\tablenotemark{c}\\
Resolution(\AA) &3.8 &3.8 & &7.5 &7.5 & 4.3 & 4.9\\
HJD (2440000+)&13343.68 & 13341.59 && 13725.64 & 13724.67 & 14003.62 & 14006.77\\
exposure time (s)&200&100 & & 300 & 100  & 900 & 900\\
S/N\tablenotemark{d}&40&95&&40 & 50 & 35 & 90 \\
parallactic angle? & yes & yes  & & yes & yes & no & no\tablenotemark{b} \\
\enddata
\tablenotetext{a}{In addition, we make use of  UVES echelle spectra obtained
on HJD 2452248.53, and cataloged in the ESO archive as 
 UVES.2001-12-05T00:47:04.840.fits and  UVES.2001-12-05T00:48:48.589.fits.}
\tablenotetext{b}{Fiber.}
\tablenotetext{c}{Dispersion is for 2-pixel binnning.}
\tablenotetext{d}{S/N per spectral resolution element measured in mid spectral range.}
\end{deluxetable}

\clearpage
\begin{deluxetable}{l c c c c c c r}
\tabletypesize{\scriptsize}
\tablewidth{0pc}
\tablenum{2}
\tablecolumns{8}
\tablecaption{\label{tab:results}  Physical Properties HV 11423}
\tablehead{
\colhead{Method}
&\colhead{$T_{\rm eff}$(K)}
&\colhead{$A_V$}
&\colhead{$\log g$}
&\colhead{$M_V$}
&\colhead{$M_K$}
&\colhead{$M_{\rm bol}$}
&\colhead{$R/R_\odot$} \\
}
\startdata
\cutinhead{2004 Dec (K0~I, V=11.84)}
Spectral fit &  4300&1.35&-0.2&-8.4&\nodata&-9.1&1060 \\
V-K            &  4480&1.35\tablenotemark{a}&-0.1&\nodata&-11.1&-9.1& 960\\
\cutinhead{2005 Dec (M4~I, V=12.46)}
Spectral fit & 3500 & 0.10 & -0.4 & -6.5 & \nodata & -8.5 & 1220 \\
V-K            & 3635 & 0.10\tablenotemark{a}&-0.2 & \nodata & -11.1&-8.3 & 1000\\
\enddata
\tablenotetext{a}{The spectral fit $A_V$ is adopted in determining the V-K physical properties.}
\end{deluxetable}

\clearpage
\begin{deluxetable}{l l l l c c}
\tabletypesize{\scriptsize}
\tablewidth{0pc}
\tablenum{3}
\tablecolumns{6}
\tablecaption{\label{tab:iras} SMC RSG IRAS Sources\tablenotemark{a}}
\tablehead{
\colhead{Name\tablenotemark{b}}
&\colhead{Type}
&\colhead{Ref}
&\colhead{IRAS Source}
&\colhead{$S_{12}$\tablenotemark{c}[Jy]}
&\colhead{$K_0-$[S12]\tablenotemark{d}} 
}
\startdata         
SMC005092  &  M1~I  &  1                        &  F00432-7321       &   0.15                   &   2.4\\
SMC018592  &  K5-M0 Ia,K0-2~I  &1,2    &   F00493-7259        &  0.17                    &  1.9\\
HV11423        & K0-1~I/M4~I & 3              &  F00592-7153        &   0.11                   &   1.9\\
SMC055188  &  M3-4~I  & 4                     &  F01014-7218\tablenotemark{e}  &         0.11       &              2.6\\
\enddata
\tablerefs{For spectral types:      
(1) Elias et al.\ 1985;
(2) Massey \& Olsen 2003;
(3) This paper
(4) Levesque et al. 2007.
}
\tablenotetext{a}{Other  red stars seen towards the SMC listed in Table 9B of
Massey (2002) that were IRAS
sources but which are not included here: [M2002] SMC035133 (IRAS F00542-7334) and
[M2002] SMC071698 (IRAS F01094-7152) have no spectral type or
radial velocity information, and the 12$\mu$ flux is much higher than for the other sources.  The
later star also has a proper motion consistent with it being a foreground halo giant, and we conclude
both of these are foreground objects.
In addition,  [M2002] SMC077678 (IRAS F01161-7359) has no spectral
type or radial velocity information, and therefore is not considered a confirmed
RSG.  [M2002] SMC084392 (IRAS F01330-7321) has a radial velocity which
demonstrated this was a foreground disk dwarf (Massey \& Olsen 2003).}
\tablenotetext{b}{``SMC" designations are from Massey 2002; i.e., [M2002] SMC005092, etc.}
\tablenotetext{c}{Typical uncertainty in $S_{12}$ is 20\%}
\tablenotetext{d}{$K_0$ is approximated from the 2MASS $K_s$ plus 0.044 (to convert
to a ``standard" Bessell \& Brett (1988) $K$ following Carpenter 2002) minus 0.03 (to correct for average extinction in $K$).}
\tablenotetext{e} {Composite source?}
\end{deluxetable}

\clearpage

\end{document}